\documentclass[aps,prb,preprint,superscriptaddress,showpacs]{revtex4}
\usepackage{graphicx}

\begin{document}

\title{Effect of physical aging on the low-frequency vibrational density of states of a glassy polymer }
\author{E. Duval}
\affiliation{Laboratoire de Physicochimie des Mat\'{e}riaux Luminescents,
Universit\'{e} Lyon I - UMR-CNRS 5620  43, boulevard du 11 Novembre 69622
Villeurbanne Cedex, France}
\author{L. Saviot}
\affiliation{Laboratoire de Recherche sur la R\'{e}activit\'{e} des Solides,
Universit\'{e} de Bourgogne and CNRS, BP 47870, 21078 Dijon Cedex, France}
\author{L. David}
\affiliation{Laboratoire des Mat\'{e}riaux Polym\`{e}res et Biomat\'{e}riaux
Universit\'{e} Lyon I - UMR-CNRS   43, Boulevard du 11 Novembre 69622
Villeurbanne Cedex, France}
\author{S. Etienne}
\affiliation{Laboratoire de Physique des Mat\'{e}riaux UMR-CNRS 7556, Ecole des
Mines 54042, Nancy Cedex, France}
\author{J. F. Jal}
\affiliation{Laboratoire de Physique de la Mati\`{e}re Condens\'{e}e et
Nanostrutures Universit\'{e} Lyon I - UMR-CNRS 5586 43, Boulevard du 11
Novembre 69622 Villeurbanne Cedex, France}
\date{\today}

\begin{abstract}

The effects of the physical aging on the vibrational density of states (VDOS)
of a polymeric glass is studied. The VDOS of a poly(methyl methacrylate) glass
at low-energy ($<15\, meV$), was determined from inelastic neutron scattering
at low-temperature for two different physical thermodynamical states. One
sample was annealed during a long time at temperature lower than Tg, and
another was quenched from a temperature higher than Tg. It was found that the
VDOS around the boson peak, relatively to the one at higher energy, decreases
with the annealing at lower temperature than Tg, i.e., with the physical aging.

\end{abstract}


\maketitle

\section{INTRODUCTION}

The effect of the physical aging on the macroscopic thermodynamical properties
of glass are well known \cite{Str78}. On the other hand, the dependence of the
structure at the nanometric scale on the physical aging is not yet clarified.
It is expected that the possible effects on the nanostructure are very small
and hardly observable by the diverse microscopy techniques. It was shown that the  properties of
vibrations, with energies in the meV spectral range, are dependent on the
nanostructure because these vibrations are extended on nanometric lengths. For
this reason, it is expected that the experimental study of the
characteristics of low-energy vibrations will inform us about the glass
nanostructure and its changes by physical aging.

It was earlier suggested  that the excess of low-energy Raman scattering, called
boson peak and the related excess of vibrational density of states (VDOS)
(called also boson peak) is due to an inhomogeneous nanostructure: more
cohesive nanodomains are separated by less cohesive zones \cite{Duv90,Mer96}.
From this model, it is expected that the contrast of bonding or of elastic
constant between cohesive domains and softer zones weakens with aging, and
that, in consequence, the vibrational density of states (VDOS) at low-energy
decreases. From previous measurements, it is not clear that the aging induces a
decrease of the VDOS at low-energy in the spectral range of the boson peak.
Isakov et al. \cite{Isa93} showed, by inelastic neutron scattering, that the
low-energy VDOS of the $As_2S_3$ glass increases after thermal quenching. This
neutron measurement confirmed the increase of the low-temperature specific heat
observed by Ahmad et al. \cite{Ahm86} after quenching in the same glass. On the
other hand, no change of the VDOS was observed after quenching in the
polybutadiene glass \cite{Zor98}.

Several experiments aimed to test the effect of thermal treatments on the VDOS
have been carried out with the poly(methyl methacrylate) (PMMA) glass. Kanai et
al. \cite{Kan96} observed an increase  of the inelastic incoherent
neutron scattering (INS) intensity at low-energy after quenching from a
temperature of $453\, K$ of an annealed sample. However, as it will be shown in
the discussion, the conditions of this experiment do not allow to conclude that
the rejuvenation (by thermal quenching) and the physical aging (by thermal
annealing at a temperature lower than the glass transition temperature, Tg)
have opposite effects on the low-energy VDOS.
New experiments aiming to test the effect of aging on the low-energy VDOS of
PMMA were carried out  recently \cite{Et02}. It was found that the aging by
annealing at $363\, K$, i.e. $10\,K$ below Tg, of a slightly cross-linked PMMA
has no effect on the VDOS determined from INS at low temperature, $T=30\, K$.
On the other hand,  by methanol-assited aging at room temperature of the same
PMMA \cite{Et01}  a significant decrease of the VDOS in the spectral range of
the boson peak was observed \cite{Et02}. These observations were in agreement
with the Raman ones. A decrease of the Raman boson peak was observed after
methanol-assisted aging of the cross-linked PMMA at $300\, K$, and none after
annealing at $363\, K$. However, the effect of the methanol-assisted aging was
much stronger with non cross-linked PMMA \cite{Et01}. Probably, the
cross-linking makes the nanostructure change much more difficult by annealing.

From the described situation, it is not yet clear that the low-energy VDOS of
non cross-linked PMMA  decreases by aging, i.e., by annealing at a temperature
slightly  lower than Tg. A decrease of the Raman boson peak at low-temperature
was observed. But the low-energy Raman scattering is proportional to the
VDOS and to the light-vibration coupling coefficient C(E), and it is possible
that this is the decrease of this coefficient which is responsible for the
decrease of the Raman boson peak and not the VDOS. To solve this problem new
measurements of inelastic neutron scattering at low-temperature were carried
out to determine directly the effect of the physical aging on the VDOS of
PMMA. The comparison of the VDOS of an aged PMMA with the one that
is rejuvenated by thermal quenching is presented in this paper. The new
neutron experimental results with others obtained earlier are discussed to
interpret the effect of aging on the glass nanostruture.

\section{EXPERIMENTAL}

The PMMA plates of clinical grade without any additives were purchased from
Goodfellow SARL (reference ME303031). The average molecular weight is close to
$10^{6}$ and the glass transition temperature $T_{g}$ is about $390\, K$, as
determined by differential scanning calorimetry (DSC). PMMA disks with a
diameter of 5 cm and a thickness of about 0.04 cm were cut for heat treatments
and inelastic neutron scattering.

The inelastic neutron spectra were recorded on the time-of-flight  instrument
IN6 at ILL, Grenoble. The energy resolution is $\delta E = 80\, \mu eV$.
The momentum transfer range extends from $Q = 0.22\, \text{\AA}^{-1}$ to $Q =
2.06\, \text{\AA}^{-1}$. The spectra were taken in the temperature range $2
- 300\, K$, using an helium cryofurnace. The temperature of $30\, K$ was
chosen to obtain the neutron inelastic scattering by the low-energy harmonic
vibrations much higher than the one  by anharmonic or relaxational motions
\cite{Mer96}. The scattering cross-sections were obtained
after the usual standard calibrations by means of the vanadium runs and the
removal of the empty-can contributions. The VDOS for harmonic modes were
obtained by taking the average of the spectra given by the different detectors,
that is the average over the range from $Q = 0.22\, \text{\AA}^{-1}$ to $Q =
2.06\, \text{\AA}^{-1}$. It was calculated through the use of an iterative
procedure described elsewhere \cite{Fon90}. The so-obtained VDOS were corrected
by the Debye-Waller factor and the multiphonon contributions.

\section {EXPERIMENTAL RESULTS}

Two PMMA samples were compared. Sample-1 was heated for 30 minutes at $413\,
K$, quenched at room temperature, annealed for 30 days at $373\, K$ and then 15
days at $363\, K$. In the view of the energy landscape, the annealing at two
different temperatures was thought to be more efficient for aging: while the
higher temperature allows the system to cross over high energy barriers, the
lower temperature allows a stabilization into deeper wells.
Sample-2 was heated for 30 minutes at $413\,
K$ and quenched at room temperature on a copper plate. The cooling rate was
estimated to be between 50 and 100 K/s.   It was controlled, by the observation
of the Raman lines, that these thermal treatments, especially the heating at
$413\, K$, did not induce depolymerization. It was checked that the intensity
of the line at $1640\, cm^{-1}$, that corresponds to the vibration localized on
the C=C bond, is  zero for both samples. It means that the concentration of MMA
monomers is very weak, and that there is no depolymerization \cite{Sur96} even
after heating at $413\, K$ for 30 min. As a consequence, Sample-1 was
physically aged and Sample-2 was rejuvenated, its thermal history  being erased
\cite{Str78}.

The effect of aging was clearly observed by DSC. A strong endothermal peak
appeared for Sample-1, and not for Sample-2, as one can see in Fig.~1.
In addition, it was determined, from the DSC curves, that the fictive
temperatures  were respectively $T_{f} = 366\, K$ and $T_{f} = 390\, K$ for
Sample-1 and Sample-2, in agreement with the thermal treatments.

\begin{figure}
\includegraphics[width=\columnwidth]{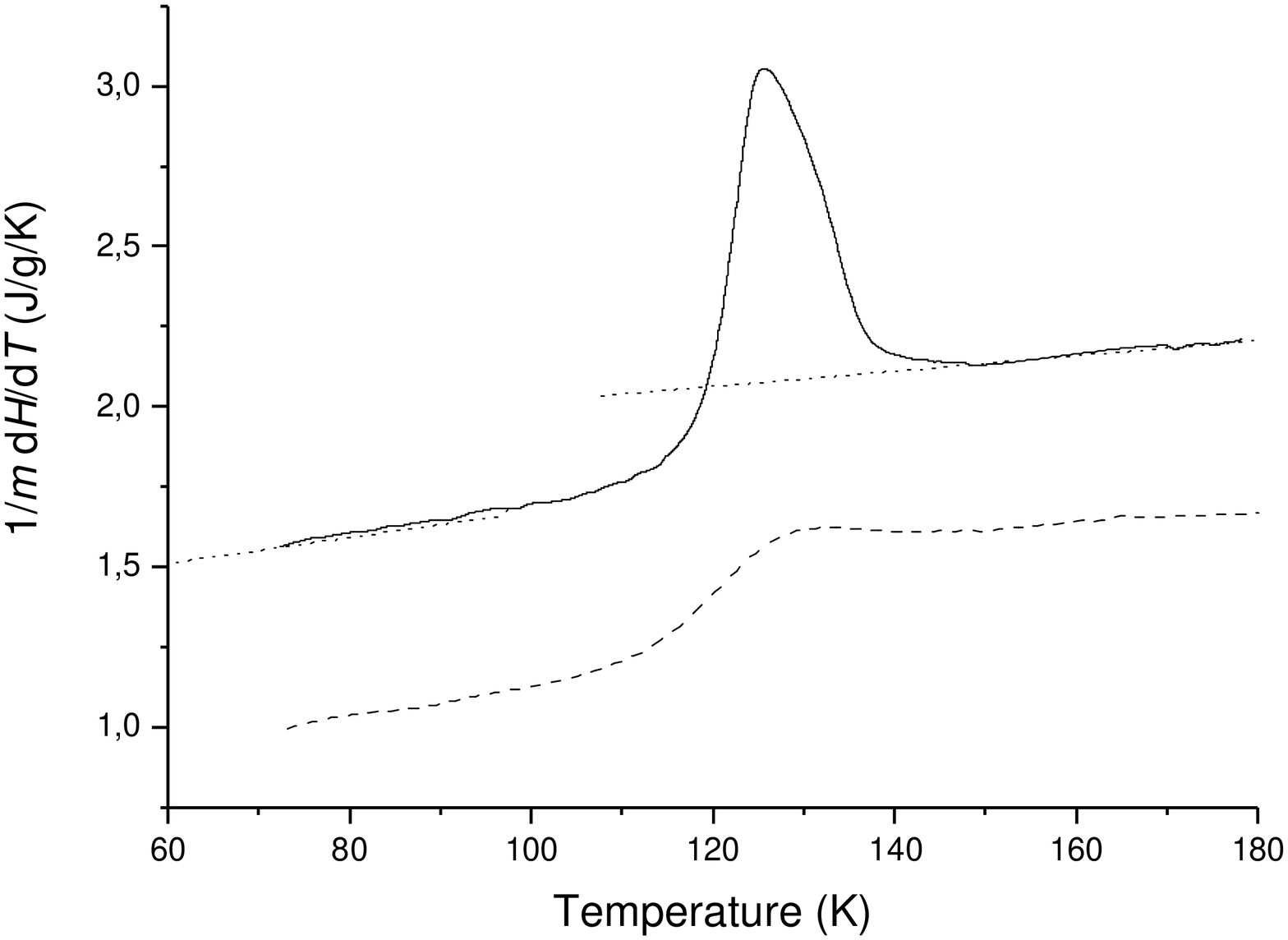}
\caption{\label{fig1}DSC curves measured at a heating rate of 20
K/min. Dotted line: reference data for $C_{p}$ values \cite{Pol89};
Full line: Sample-1 (aged PMMA); Dashed
line: Sample-2 (quenched PMMA). The data represented by the dashed line are
vertically shifted by -0.5 J/g/K for clarity.} \end{figure}

The incoherent scattering functions of both samples, $S(E)$, taken at $T = 30\,
K$, are compared in Fig.~2. They were determined by taking the average of the
spectra given by the different detectors over the range from $Q =
0.22\, \text{\AA}^{-1}$ to $Q = 2.06\, \text{\AA}^{-1}$. They are corrected
for the elastic peak, that was obtained by measuring the inelastic scattering
from the samples at $T = 2\, K$. As their shapes are identical from the
energy equal to 4 meV, they were normalized by
coincidence of the curves from the energy of 4 meV for a better comparison.
Considering the experimental error bars due to neutron counting, which
are shown on the curves at 1.25 meV, it is clear that $S(E)$ is
more intense in the $0.6 - 2\, meV$ spectral range for Sample-2 than for
Sample-1.

\begin{figure}
\includegraphics[width=\columnwidth]{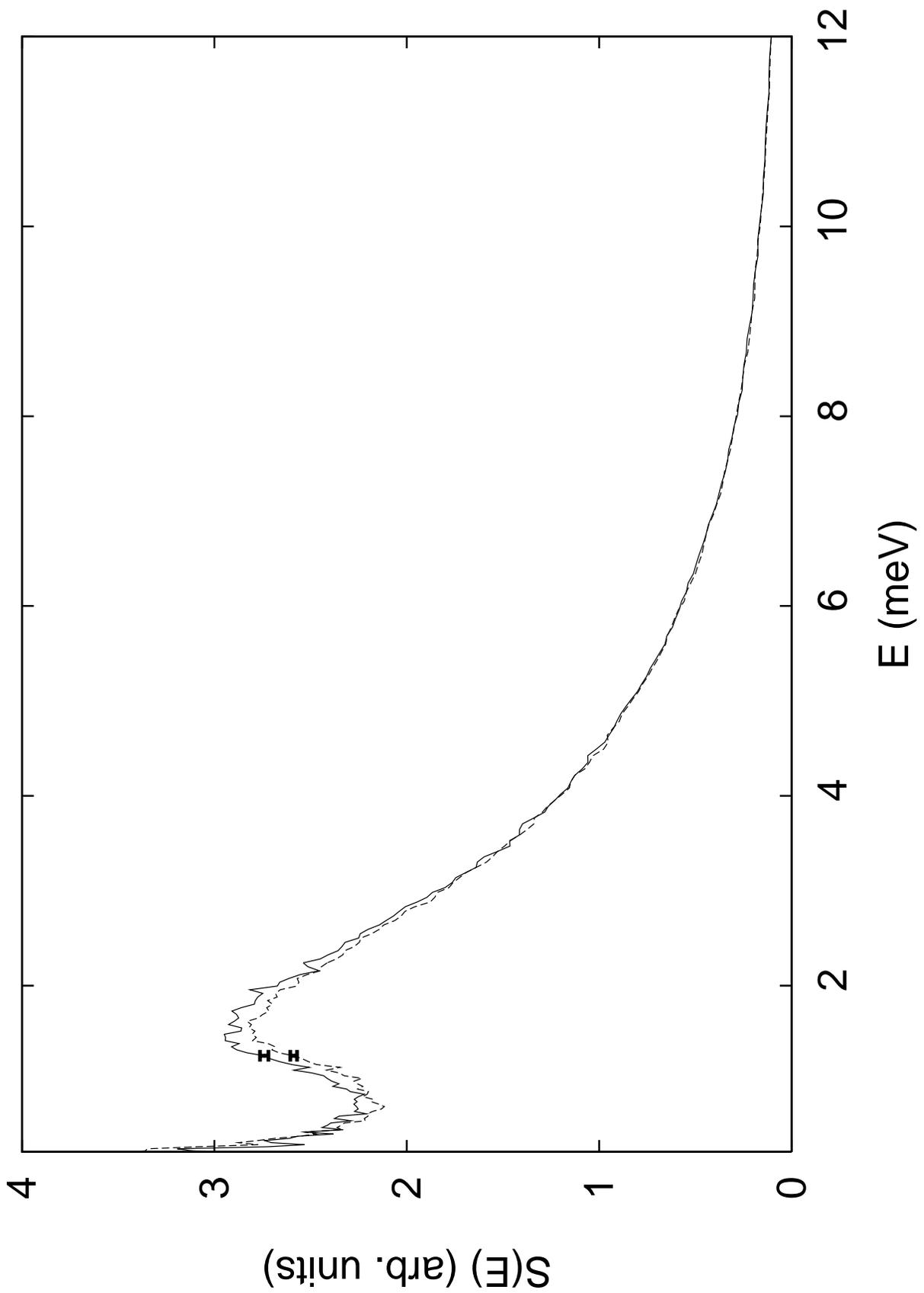}
\caption{\label{fig2}Incoherent neutron scattering function $S(E)$. Full
line: quenched PMMA; Dotted line: aged PMMA. The error bars arising from the
neutron counting are shown at 1.25 meV on both curves. } \end{figure}

The VDOS $g(E)$ were deduced from $S(E)$ \cite{Fon90}. These
VDOS divided by the square of energy, $g(E)/E^{2}$, are plotted in Figure-3. As
expected from $S(E)$ (Figure-2), the boson peak at 1.75 meV is more intense for
the rejuvenated Sample-2 than for the aged Sample-1, even if the difference is
not huge. Furthermore, it is  evident, from Figure-3, that the observed
difference originates from the VDOS for harmonic vibration modes and not for
anharmonic ones, because it does not decay monotonously from E=0 to higher
energy as expected for anharmonic motions: it is observed (inset of Figure-3)
that this difference presents a maximum around E = 1.05 meV. It was detected
that it is not the case for a sample slightly depolymerized by  heating at 413
K for a much longer time and containing some monomers: even at $T = 30 K$
neutron scattering by anharmonic motions due to the presence of monomers was
observed, in agreement with previous Raman measurements \cite{Nik96}. The VDOS
were also determined at $T = 300 K$. However, at this temperature, the neutron
scattering by anharmonic motions is relatively high for both samples and masks
the scattering by harmonic vibrations at low-energy.

\begin{figure}
\includegraphics[width=\columnwidth]{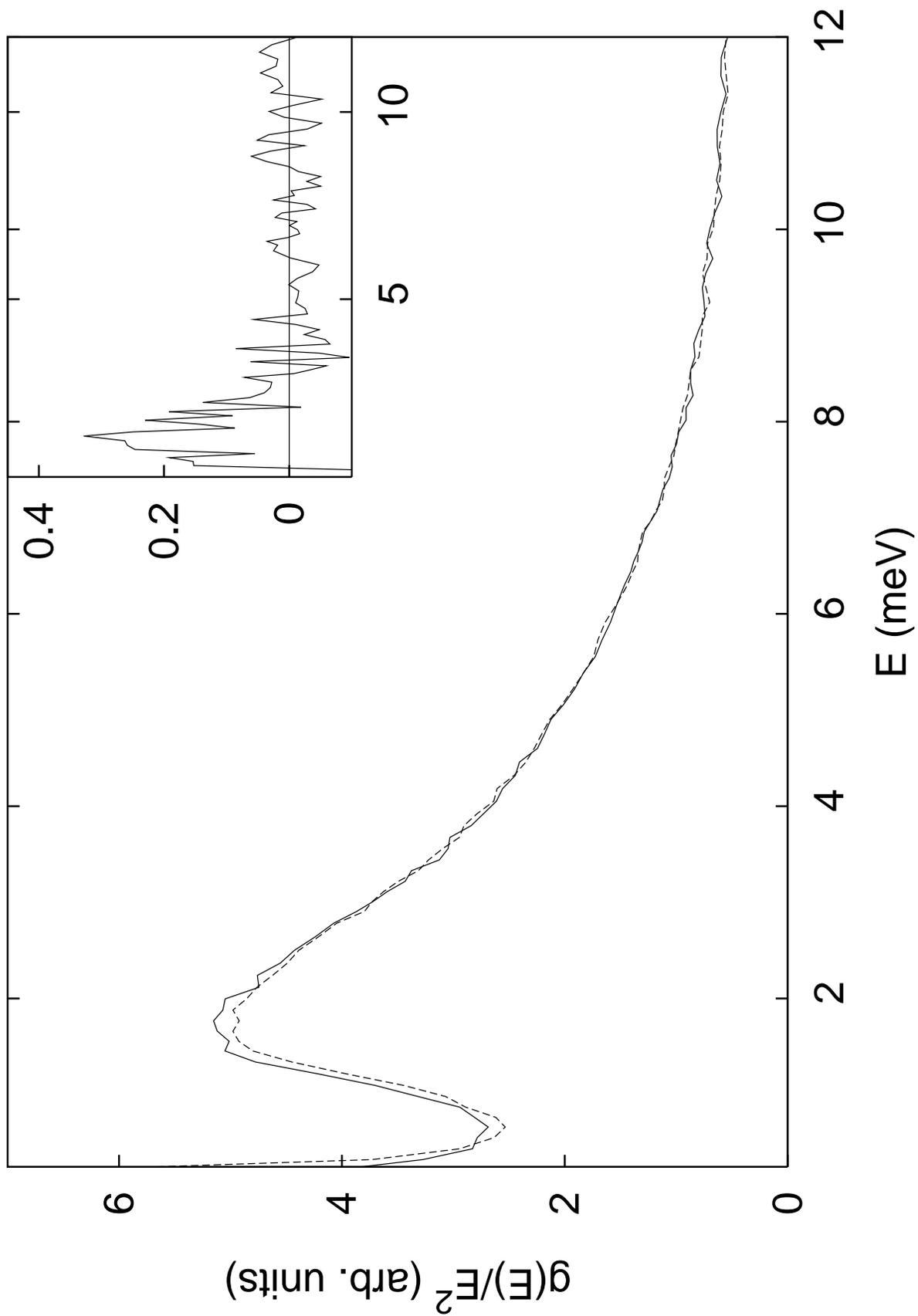}
\caption{\label{fig3}Vibrational density of states, $g(E)/E^{2}$. Full
line: quenched PMMA; Dotted line: aged PMMA. The inset shows the
difference curve: $[g(E)/E^{2}]_{quenched}-[g(E)/E^{2}]_{aged}$}
\end{figure}

\section {DISCUSSION}

In this described experiment much care was taken, on the one hand, to make sure
that the heat treatments induce only physical aging and not a chemical one like
depolymerization, and, on the other hand, to obtain the effect of aging on the
VDOS corresponding to harmonic vibration modes. In a previous study of the
effect of annealing on the low-energy excitations in PMMA, the conditions of
the experiment were different  \cite{Kan96}. Indeed, these authors of this
study compared a sample annealed 7 days at $T = 363\, K$ to a sample quenched
from $T = 453\, K$. It was shown that at this temperature depolymerization
occurs and induces an increase of the monomer concentration \cite{Sur96}. The
reversible effect was observed by subsequent annealing at a temperature lower
than Tg \cite{Sur96}. As a consequence, by such a thermal treatment
\cite{Kan96}, this is a partially chemical aging which is obtained instead of a
pure physical aging. Furthermore, in this experiment \cite{Kan96} the inelastic
scattering was measured at $T = 73\, K$. At this temperature the scattering by
anharmonic or relaxational motions is not negligible \cite{Mer96}. In these
conditions and after observation of the published spectra, it is likely that
the difference in the inelastic neutron scattering of annealed and quenched
PMMA observed by Kanaya et al. is mainly due to anharmonic motions.

In another glassforming polymer, polybutadiene, no effect of annealing was
observed on the VDOS by measuring the inelastic neutron scattering
\cite{Zor98}. It could be suggested that the decrease of the low-energy VDOS by
aging is not general for glass-forming polymers. However, the used thermal
treatments can explain the non-observation of difference in the VDOS between
annealed and quenched polybutadiene. The aged sample was heated for 36 h at $T
=169\, K$ (11 K below Tg), and the quenched sample was heated for 1 h at 270 K
and cooled at 100 K with a rate of 2 K/min. The cooling rate used for this
quenching was possibly too slow to avoid an undesidered effect of aging during
the cooling from 270 K to 100 K.

In a very recent study of a crystallisable silicate material by inelastic
neutron scattering, it was observed that the intensity of the boson peak of a
hyperquenched ($10^{6}\, K/s$) sample decreased strongly after annealing at a
temperature below Tg \cite{Ang03}. The neutron measurements were performed at
room temperature, and probably the measured VDOS  is partially accounted for
anharmonic vibrational modes. A similar behavior was earlier observed with a
metallic glass \cite{Suck89}. These results were discussed in the frame of the
potential energy landscape model \cite{Ang03}.

A simple explanation  of the decrease of the VDOS by aging, in our experiment,
could be the increase of sound velocity concomitant with the increase of
density, and in consequence a decrease of the Debye VDOS contribution
\cite{Isa93}. However, the  difference of $g(E)/E^{2}$  between aged and
rejuvenated PMMA is expected to be constant as a function of E. It is not
the case, this difference has a maximum around 1.05 meV (inset of Figure-3).
Instead of having an effect on the Debye VDOS, the aging would change the
excess of VDOS responsible for the boson peak. The interpretation of the
low-frequency modes in the boson peak is still a matter of
debate. Several theoretical models are based on the system dynamics without
consideration of the structure, among which the very recent landscape model
\cite{Gri03}. Here, we will consider the nanostructural point of view, which is
not in contradiction with the dynamical approach and may be relevant to
discuss the aging effect. The boson peak was previously
\cite{Duv90,Mer96} interpreted by an inhomogeneous glass elasticity or cohesion
at the nanometric scale: more cohesive domains would be separated by softer
zones. This view is supported by a recent simulation of the elasticity of small
amorphous systems \cite{Wit02}. Previous experimental results are in agreement
with quasi-periodic arrangements of the cohesive domains on short distances
\cite{Mer98,Duv98}, so that the boson peak would correspond to a Van Hove
singularity associated to the quasi-periodic arrangements \cite{Duv02}. Such
interpretation can be compared to the model of Taraskin et al. \cite{Tar01}, in
which the boson peak is related to the lowest Van Hove singularity of the
reference crystalline system pushed down by the effect of disorder-induced
level repulsion. In the frame of the inhomogeneous glass nanoelasticity model,
the decrease of the low-energy VDOS by aging can originate from a decrease of
the contrast in the elasticity at the nanometric scale. This would be in
agreement with the observations of Isakov et al. about the effect of aging in
the $As_{2}S_{3}$ glass \cite{Isa93}. These authors determined that the
relative effect of aging on the low-energy VDOS was the same as the one on the
light-vibration coupling coefficient $C(E)$ in Raman scattering. This
coefficient $C(E)$ increases with the amplitude of the dielectric
susceptibility fluctuations \cite{Jac89}. The dielectric susceptibility is
directly related to the elastic constants by the elasto-optic tensor.
Therefore, the dielectric susceptibility contrast at the nanoscale goes like
the elastic constant one, and one may conclude that the decrease of $g(E)$ and
$C(E)$ have a common origin: the smoothing of the elastic constant contrast at
the nanometric scale. The same conclusion can be proposed for PMMA.
Unfortunately, it was not possible to perform a quantitative comparison between
$g(E)$ and $C(E)$ in the case of PMMA, because the effect of aging on these
quantities are relatively much weaker for this glass than for $As_{2}S_{3}$
\cite{Ahm86,Isa93}.

The results obtained by the inelastic neutron scattering study presented in
this paper must be compared to other previous ones. Same measurements were
carried out with PMMA slightly cross-linked by trimethyl-1,1,1-propane
trimethacrylate (TRIM) \cite{Et02}. With this cross-linked PMMA, no effect of
aging was observed on the low-energy Raman scattering by harmonic vibration
modes. It is believed that it is due the presence in the zones between domains
of cross-links which reinforce the linking  of a domain with its next-neighbors
and, then, precludes the decrease of the contrast between soft zones
and cohesive domains  with aging. Methanol-assisted aging of PMMA at
room temperature was also studied \cite{Et01,Et02}. PMMA plates was soaked for
more than one month in methanol at room temperature and then dried. By
mechanical relaxation measurement, it was shown that the effect of this
treatment corresponds to an aging at room temperature accelerated by the
presence of methanol. This can be explained by a preferential location of the
methanol in the softest zones, which makes easier the molecular segmental
motions.

It may be noticed that the tendency towards a more homogeneous structure by
aging was deduced from elastic light scattering by
Takahara et al. in the case of PMMA \cite{Tak99}. These authors determined the correlation length
and the amplitude of density fluctuations at the scale of 200 nm. They found
that the values of these two fluctuation characteristics decreased after an
aging at $T = 353\, K$ for less than 10 hours. These conclusions are not
exactly comparable to the ones suggested in this paper. In the present work,
the explored length scale is 10 times smaller, and deals with the fluctuations
of elastic constant, and not with the density ones.

The effect of aging or rejuvenation by thermal treatments on the low-energy
vibrations, is now compared to that of a plastic deformation
\cite{Merm96}. It is well-known that a plastic deformation rejuvenates a
glass \cite{Str78,Has93,Hasa93,Hod95,Utz00}. The effect of plastic
deformation by shearing on the low-energy vibrations was investigated by
Raman scattering in the case of PMMA and polycarbonate (BPA-PC) glasses
\cite{Merm96}. An increase of the Raman scattering
intensity in the region of the boson peak was observed. This increase of Raman
intensity can be attributed, at least partially, to an increase of VDOS.
Furthermore, as a confirmation, no increase of low-energy Raman scattering
was observed after plastic deformation of the cross-linked PMMA \cite{Merm96},
like after thermal quenching \cite{Et02}. These results confirm the effect of
aging (and rejuvenation) on the low-energy VDOS and on the glass nanostructure,
consistently with the proposed interpretation.

It is noted that the experimental results and their interpretation are
in agreement with, on the one hand, the view of a vitrifying liquid, that was
deduced by Stillinger \cite{Stil88} from the potential energy landscape, "as a
patchwork of relatively strongly bonded molecular domains separated by
irregular walls of weakened bonds"; and, on the other hand, with the recent
numerical simulation of aging and rejuvenation, which shows that the aging
induced a pronounced decrease in the system pressure, and that local
compression and expansion of rearranging clusters of atoms appear as the shear
deformation proceeds and, therefore, erases the aging \cite{Utz00}. This
description is not far from the one involving the arrangement of more cohesive
domains separated by less cohesive zones as proposed earlier
\cite{Duv90,Mer96,Mer98,Duv98}, the contrast between cohesive domains and
softer zones decreasing with aging, and inversely increasing with rejuvenation,
as suggested in this discussion.

Finally, it is remarked that the observed decrease  of the VDOS with aging
confirms qualitatively the very recent predictions deduced from
the energy landscape model \cite{Gri03}.

\section{CONCLUSION}

The inelastic neutron scattering experiment described in this paper shows
clearly a lower harmonic vibration density of states for aged poly(methyl
methacrylate) than for thermally quenched (rejuvenated) one. Much care was
taken to avoid chemical effects by thermal treatment. The neutron
measurements were carried out at low-temperature to obtain the scattering by
harmonic vibrations. The effect of thermal quenching is compared to the one of
a plastic shearing observed earlier. The rejuvenations obtained  by both
thermal and mechanical treatments  are characterized by an increase of the
low-energy vibrational density of states. These effects on the vibrational
density of states are in agreement with the model of an inhomogeneous cohesion
in the polymeric glasses at the nanometric scale, and with the energy landscape
deductions. It would be interesting to establish a relation between the glass
nanostructure or the nanoelasticity and the profile of the energy landscape.

\begin{acknowledgments}
The authors are grateful to J. Ollivier from ILL for his assistance in the
neutron measurements, and to A. Mermet for discussion. They thank warmly C. A.
Angell to send them a copy of his paper \cite{Ang03} before publication and for
his comments.
\end{acknowledgments}

\bibliography{publis}
\end{document}